\documentclass[twocolumn,showpacs,preprintnumbers,amsmath,amssymb]{revtex4}
\usepackage[english]{babel}
\usepackage[latin2]{inputenc}
\usepackage{epsfig}

\begin{document}
\preprint{M\v{s}/2}

\title{Lattice constant in diluted magnetic semiconductors
(Ga,Mn)As}

\author{J.~Ma\v{s}ek, J.~Kudrnovsk\'{y}, and F.~M\'{a}ca}

\affiliation{Institute of Physics, AS CR, Na Slovance 2, CZ--182
21 Prague 8, Czech Republic}

\date{February 7, 2002}

\begin{abstract}
We use the density-functional calculations to investigate the
compositional dependence of the lattice constant of (Ga,Mn)As
containing various native defects. The lattice constant of perfect
mixed crystals does not depend much on the concentration of Mn.
The lattice parameter increases if some Mn atoms occupy
interstitial positions. The same happens if As antisite defects
are present. A quantitative agreement with the observed
compositional dependence is obtained for materials close to a
complete compensation due to these two donors. The increase of the
lattice constant of (Ga,Mn)As is correlated with the degree of
compensation: the materials with low compensation should have
lattice constants close to the lattice constant of GaAs crystal.
\end{abstract}

\pacs{71.15.Ap, 71.20.Nr, 71.55.Eq}

\maketitle

%
%

Diluted magnetic III-V semiconductors (DMS), such as
Ga$_{1-x}$Mn$_{x}$As, combine semiconducting and ferromagnetic
properties \cite{Ohno98,Dietl00,Twardowski00,Konig02} and are
attractive for applications in spin electronics. These materials
have been extensively studied in the last years, both
experimentally and theoretically.

There is, however, still not much known about the details of the
crystal structure of these materials and about the incorporation
of Mn atoms. It is generally believed that in well defined samples
the volume of the MnAs precipitates is reduced to zero, and that
Mn simply substitutes for the host cation in a tetrahedral
(zinc-blende or wurtzite) crystal structure. Only recently it was
suggested \cite{Masek01,Maca02} and experimentally proved
\cite{Yu02} that a portion of Mn occupies interstitial rather than
substitutional positions in the zinc-blende lattice of (Ga,Mn)As.
The interstitial Mn atoms act as double donors
\cite{Masek01,Maca02,Sanvito02,Erwin02}, in contrast to Mn atoms
in the substitutional positions that are known to be acceptors.

Almost unnoticed remains the surprising fact that the lattice
constant of (Ga,Mn)As increases with increasing concentration of
Mn \cite{Ohno99}. According to the atomic radii \cite{Langes}, Mn
atoms are smaller ($R_{\rm Mn}$ = 1.17~\AA) than Ga atoms ($R_{\rm
Ga}$ = 1.25~\AA) and, in the simplest approximation, the lattice
constant should be expected to decrease rather than to increase.
This is also a result of a recent theoretical study \cite{Zhao02}
of the structure of zinc-blende $\alpha$-MnAs. According to these
calculations the lattice constant of $\alpha$-MnAs is smaller then
the lattice constant of GaAs.

On the other hand, the lattice constant of GaAs is well known to
increase in the presence of As antisite defects
\cite{Liu95,Staab01}. The MBE-grown GaAs crystals may contain up
to 1 atomic percent of these defects and a large amount of the
antisite defects is expected also in (Ga,Mn)As \cite{Dietl01}.
Being donors, they have an important role in the compensation of
Mn acceptors. It was also shown recently \cite{Masek02} that
formation energy of an As antisite defect in (Ga,Mn)As decreases
remarkably with the increasing content of Mn and that the
concentration of As antisites should be correlated with the
concentration of Mn atoms. This indirect mechanism, i.e., the
increasing number of the As antisites due to the addition of Mn,
could be a possible explanation of the observed increase of the
lattice constant of (Ga,Mn)As.

Also the presence of the interstitial Mn atoms can be the reason
for the observed expansion of the lattice \cite{Maca02}, assuming
only that the number of the interstitials increases proportionally
to the total concentration of Mn.

In this paper, we put these intuitive considerations on serious
grounds by using the density-functional calculations. We consider
GaAs crystal with small but finite concentration of various
impurities, such as Mn atoms in either substitutional or
interstitial positions and As atoms in the cationic sublattice. We
use a tight-binding linear muffin-tin orbital (TB-LMTO) method to
describe the electronic structure of these imperfect crystals. The
charge self-consistency is treated in the framework of the local
spin-density approximation with the Vosko-Wilk-Nusair
parametrization \cite{Vosko80} for the exchange-correlation
potential. The crystal potential is considered within the
atomic-sphere approximation (ASA) with empty spheres in
tetrahedral interstitial positions for a good space filling.

The substitutional disorder due to the random distribution of
either Mn or As atoms on the cationic sublattice as well as the
random distribution of Mn atoms in the interstitial positions is
treated in the coherent-potential approximation (CPA) -- for
details see \cite{Turek97}. The advantage of the CPA is that it is
well suited for systems with low concentrations of impurities,
assuming the unperturbed, zinc-blende symmetry of the mixed
crystals. The CPA treatment, on the other hand, neglects the
relaxation of the lattice around the impurities.

The lattice constant $a$ is used as a variable parameter and the
total energy is calculated for approximately 10 values of $a$
around the calculated lattice constant $5.57~{\rm \AA}$ of the
pure GaAs. The minimum of the total energy from the
density-functional calculations with respect to $a$ is found by
using a cubic interpolation scheme.

%
%
\begin{figure}
\epsfig{file=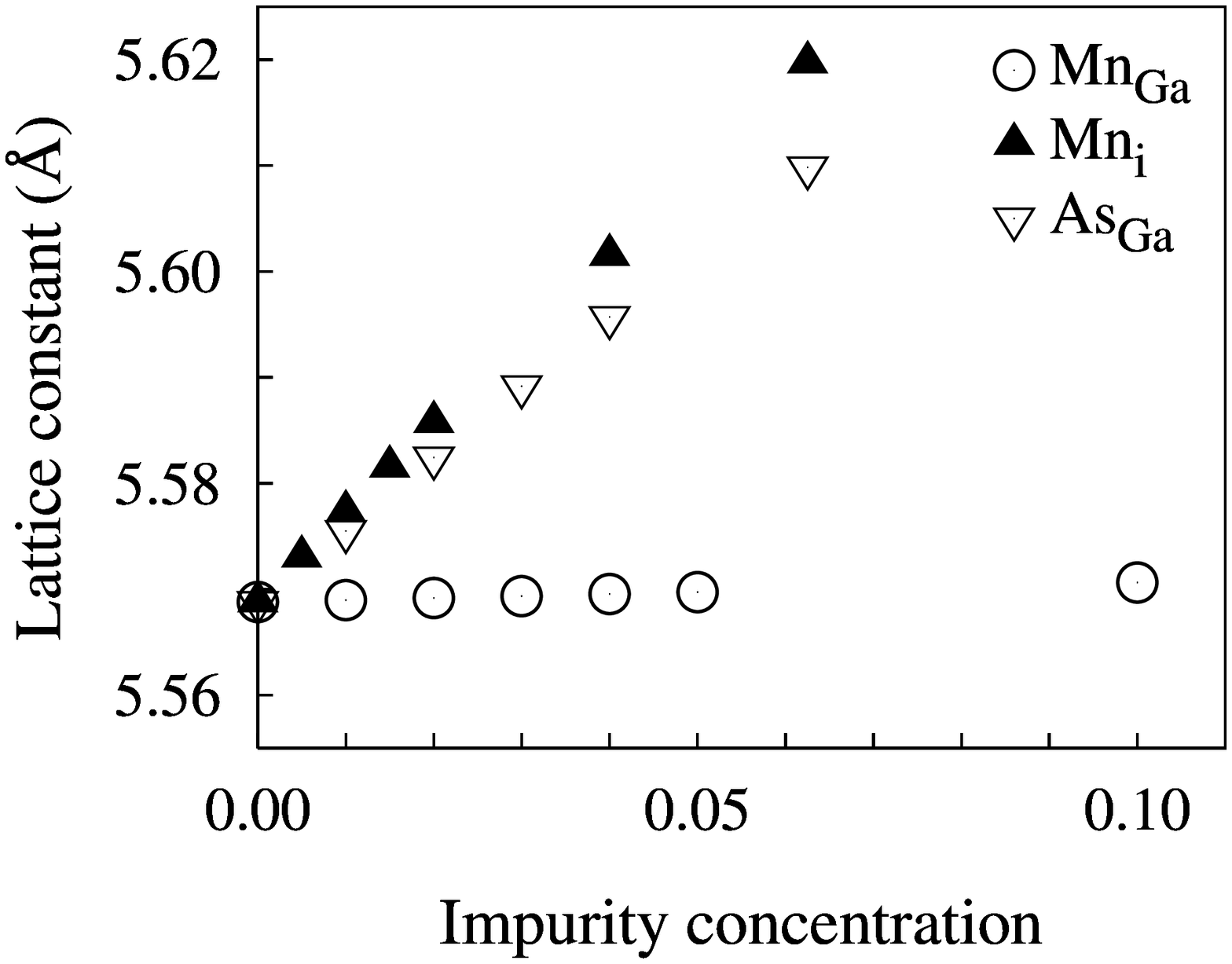,height=8cm,width=5.5cm,angle=270}
\caption{\label{fig:epsart} Calculated lattice constant as a
function of the concentration of the impurities:\\ (a) Mn atoms in
the substitutional positions (circles),\\ (b) Mn atoms in the
interstitial positions (full triangles),\\ (c) As antisite defects
(empty triangles).}
\end{figure}

We start with ideal mixed crystals Ga$_{1-x}$Mn$_{x}$As where all
Mn atoms are in the substitutional positions. We calculated the
lattice constants for a series of materials with $x$ = 0.01, 0.02,
0.03, 0.04, and 0.05, and also for $x$ = 0.10. The calculated
dependence of the lattice constant on the concentration is linear
up to $x = 0.10$,
\begin{equation}
a(x) = a_{0} + 0.02 x {\rm ~~(\AA)}.
\end{equation}
It is in a good agreement with the Vegard law. The calculated
lattice constant of GaAs crystal, $a_{0} = 5.569~{\rm \AA}$, is
smaller than the observed value $a_{0}^{\rm exp} = 5.653~{\rm
\AA}$ \cite{DST}. This is a result of the local spin-density
approximation (LSDA) combined with ASA. In our study, however, we
are not primarily interested in the absolute values of the lattice
constant but in its variation due to the changes of the chemical
composition. We assume that this systematic underestimation has
only a minor effect on the compositional dependence of the lattice
constant $a$, characterized by the linear coefficient in (1). This
coefficient is very small. Even at the highest concentration of Mn
atoms ($x$ = 0.1), the lattice constant of the mixed crystal $a =
5.571~{\rm \AA}$ does not differ from the lattice constant of the
pure GaAs by more than 0.05 percent.

This means that the calculated changes of $a(x)$ are by an order
of magnitude smaller than the observed values \cite{Ohno99}. This
result is in a good correspondence with the recent finding that
the local relaxations around the substitutional Mn impurity are
very small and have only a small impact on its electronic
configuration \cite{Mirbt02}. According to Zhao et al.
\cite{Zhao02}, the lattice constant of Ga$_{1-x}$Mn$_{x}$As may
even decrease with increasing $x$. From this point of view, we
conclude that the substitution of Mn atoms into the cationic
sublattice has a negligible effect on the lattice constant of
Ga$_{1-x}$Mn$_{x}$As and that the observed expansion of the
lattice of the (Ga,Mn)As mixed crystals should be ascribed to
other lattice defects.

To estimate the effect due to the interstitial Mn atoms, we
consider first hypothetical materials GaMn$_{z}$As in which all Mn
atoms are in the interstitial positions. We assumed only the
interstitial positions surrounded by As atoms (cf. \cite{Masek01})
that are thermodynamically more favorable than the positions with
neighboring Ga atoms. As shown in Fig. 1, the calculated lattice
constants of GaMn$_{z}$As with $z$ = 0.01, 0.02, 0.03, and 0.04
lie on a straight line
\begin{equation}
a(z) = a_{0} + 0.86 z {\rm ~~(\AA)}.
\end{equation}
The change of the lattice constant due to the addition of Mn atoms
is in this case much stronger as compared to the case of
substitution. Assuming for simplicity that the increase of the
lattice constant is only due to the presence of the interstitial
Mn atoms, we can use (2) to estimate the number of these defects
in the material. The experimental compositional dependence of the
lattice constant of (Ga,Mn)As is \cite{Ohno99}
\begin{equation}
a(\tilde{x}) = a_{0}^{\rm exp} + 0.32 \tilde{x} {\rm ~~(\AA)},
\end{equation}
where $\tilde{x}$ denotes the total concentration of Mn atoms
($\tilde{x} = x + z$). Combining Eqs. (1), (2), and (3) we get $z
\approx \tilde{x} / 3$. This result is close to the experimental
finding \cite{Yu02} and to a simple estimate of Ref.
\cite{Maca02}. It is important to point out that materials with
such a ratio between acceptors and donors are almost completely
compensated, with a strongly reduced doping efficiency of Mn.

It should be noted that the hypothetical materials used to obtain
Eq. (2) contain no substitutional Mn and are n-type
semiconductors. This is the reason why we have performed
additional calculations also for p-type materials containing 5
atomic percent of the substitutional Mn. We consider the
$z$--dependence of the lattice constant for
Ga$_{0.95}$Mn$_{0.05}$Mn$_{z}$As mixed crystals. In this case we
obtain that the linear coefficient in $a(z)$ is 1.26 instead of
0.86 as in Eq. (2). The real concentrations of the interstitial Mn
atoms are, however, small ($z < 0.01$) so that both values of the
linear coefficient describe the increase of the lattice constant
with a reasonable accuracy. In the following, we shall consider a
simple modification of Eq. (2) with the average value 1.05 of the
linear coefficient instead of introducing corrections proportional
to the product $x z$.

\begin{figure}
\epsfig{file=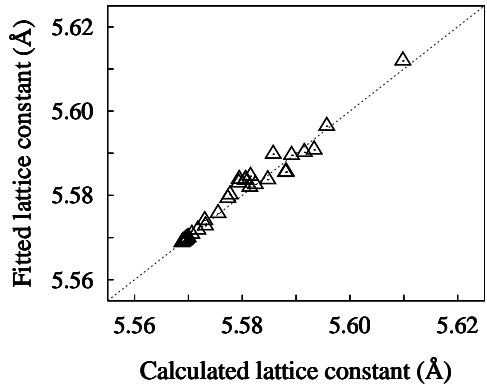,height=7cm,width=5.5cm,angle=270}
\caption{Correlation plot for he lattice constant. The values
obtained according to Eq. (5) (y-axis) are plotted against the
results of the density-functional calculations (x-axis).}
\end{figure}

The dependence of the lattice constant on the concentration of As
antisite defects was treated in the same way. We considered a
series of hypothetical non-stoichiometric crystals
Ga$_{1-y}$As$_{y}$As with a random distribution of the additional
As atoms on the cationic sublattice. In Fig. 1 we show that the
lattice constant increases with the increasing concentration $y$
of the antisite defects. The data for $y$ = 0.01, 0.02, 0.03, and
0.04 can be fitted by a linear function
\begin{equation}
a(y) = a_{0} + 0.69 y {\rm ~~(\AA)}.
\end{equation}
The substitution of the As atoms into the Ga sublattice has a much
stronger effect on the lattice expansion as compared to the
substitution of Mn. The value of the linear coefficient compares
well with the lattice expansion of GaAs due to As antisite defects
obtained recently using the the large-unit-cell (LUC) calculations
\cite{Staab01}. The coefficient in the experimental function (3)
is approximately one half of the linear coefficient in (4). This
means that the number of the antisite defects necessary to explain
the observed increase of the lattice constant is, with a good
accuracy, proportional to the total concentration of Mn, namely $y
\approx \tilde{x} / 2$. This result is not surprising because it
implies an almost complete compensation, which is actually
observed in (Ga,Mn)As.

Finally, Eqs. (1), (4), and a modified Eq. (2) can be summarized
to a simple formula for the compositional dependence of the
lattice constant of (Ga,Mn)As,
\begin{equation}
a(x,y,z) = a_{0} + 0.02 x + 0.69 y + 1.05 z {\rm ~~(\AA)}.
\end{equation}
The additivity of three contributions on the right--hand side of
(5) was checked by calculations of the lattice constant for
several compositions with different values of concentrations $x$,
$y$, and $z$. The validity of the formula (5) is illustrated in
Fig. 2 in which we present a correlation plot of the values of
$a(x,y,z)$ as obtained from Eq. (5) against the results of the
density--functional calculations. Most of the points are found
close to the diagonal. This means that Eq. (5) is applicable with
a reasonable accuracy to the whole low--concentration range of the
mixed (Ga,Mn)As crystals.

%
%
The observed dependence of the lattice constant can be obtained
assuming that the concentration of either interstitial Mn atoms or
As antisites increases proportionally to the nominal concentration
of Mn. A rough estimate of the proportionality coefficients shows
that in both cases the number of the native defects is such that
the system is highly compensated.

Combining the calculated linear coefficients in Eq. (5) with the
condition $x + z = \tilde{x}$ and with the expression
\begin{equation}
\eta \tilde{x} = x - 2 y - 2 z
\end{equation}
for the doping efficiency $\eta$ we can speculate about the values
of $x$, $y$, and $z$. For a realistic degree of compensation, $0.1
\leq \eta \leq 0.2$, the fit of (3) to (5) does not result to a
preferential occurrence of either Mn interstitials or As
antisites. This result indicates that both donors are equally
important for the compensation in (Ga,Mn)As. It is also in a good
correspondence with the fact, that the formation energies of both
Mn interstitials \cite{Erwin02} and As antisites \cite{Northrup93}
have roughly the same value ($\approx$ 2~eV).

The dependence of the total energy $E_{tot}(a)$ on the lattice
constant $a$ can also be used to determine the elastic modulus
\begin{equation}
B = \frac{1}{9a} \frac{d^{2}E_{tot}(a)}{da^{2}}
\end{equation}
for the (Ga,Mn)As mixed crystals and its compositional dependence.
Our results indicate that the bulk modulus $B$ does not depend
much on the concentration of substitutional Mn atoms, as shown in
Fig. 3. It decreases in the presence of the interstitial Mn atoms
and in particular in the presence of the As antisites. The
softening of the lattice is not surprising because these defects
disturb the crystal bonding. It should be noticed, however, that
the elastic modulus, as compared to the lattice constant, is much
more sensitive both to the (neglected) lattice relaxation and to
the detailed shape of the potential.

\begin{figure}[h!]
\epsfig{file=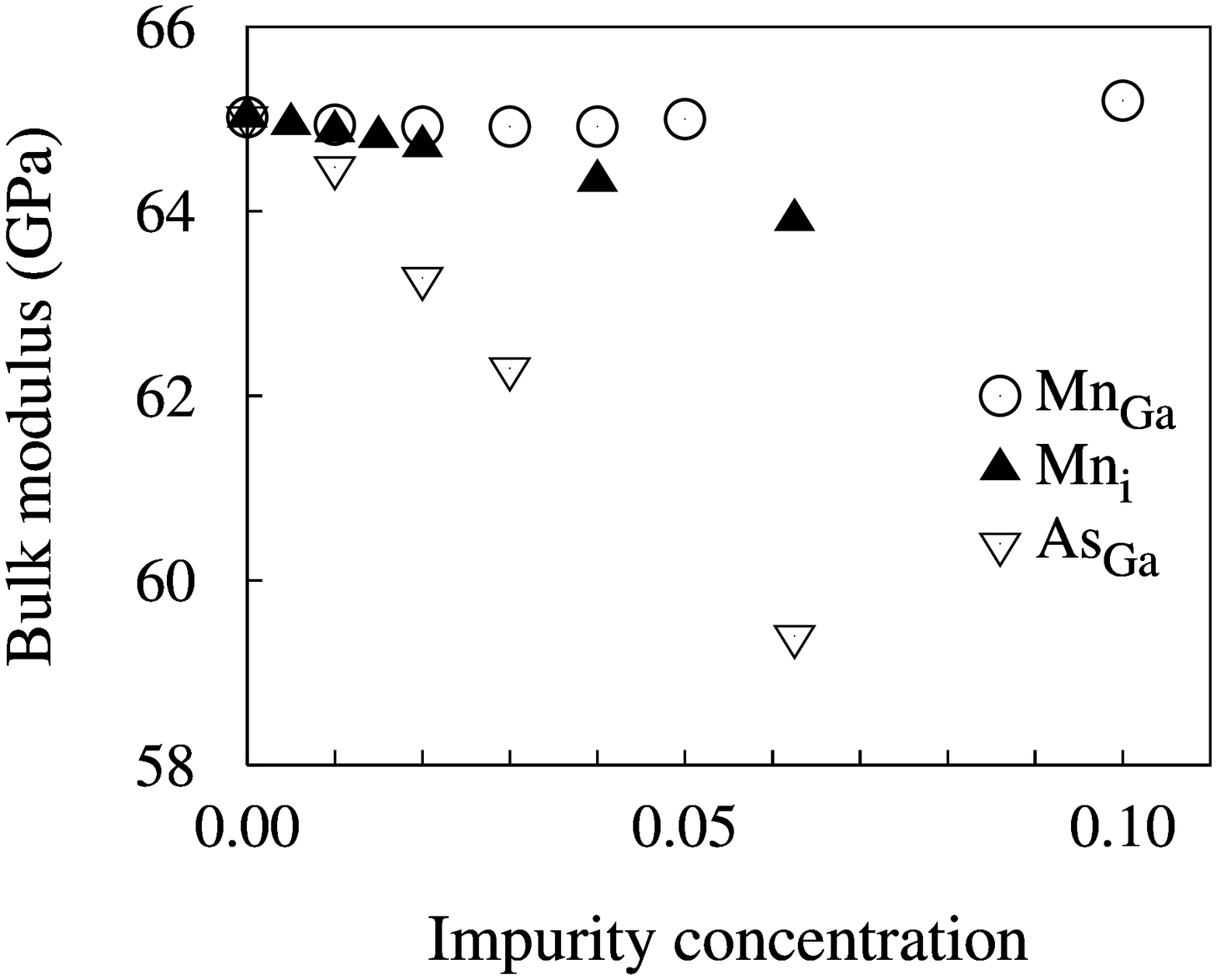,height=8cm,width=5.5cm,angle=270}
\caption{Calculated bulk modulus as a function of the
concentration of the impurities:\\ (a) Mn atoms in the
substitutional positions (circles),\\ (b) Mn atoms in the
interstitial positions (full triangles),\\ (c) As antisite defects
(empty triangles).}
\end{figure}

We also performed complementary full-potential LAPW \cite{WIEN97}
calculations of the lattice constants for supercells
Ga$_{15}$MnAs$_{16}$ and Ga$_{16}$MnAs$_{16}$ with Mn in either
substitutional or interstitial positions. Both LSDA and GGA
versions of the density functional were used. The results of the
supercell calculations confirm the basic finding of the CPA study
that the substitution of Mn in the cationic lattice has a
negligible effect on the lattice constant. In the case of the
interstitial Mn, remarkable increase of the lattice constant is
found, similar to Eq. (2). In this case, the preliminary
calculations also indicate, in contrary to the case of
substitution \cite{Mirbt02}, the importance of the lattice
relaxations around the Mn impurity. The detailed discussion of the
Mn interstitials in (Ga,Mn)As will be given elsewhere.

We conclude that the lattice constant can be used as a simple
indication of the quality of the diluted (Ga,Mn)As mixed crystal.
The lattice constant increasing with the content of Mn seems to be
an inherent property of materials with a large number of
Mn-induced native defects. These defects lower the doping
efficiency of Mn in the mixed (Ga,Mn)As crystals and, in turn,
also the Curie temperature. On the other hand, the desirable
samples with low concentrations of compensating donors and with
most of Mn atoms substituted into the cationic sublattice are
expected to have almost the same lattice constant as the
underlying GaAs crystal.

\subsection* {Acknowledgment}
This work has been done within the project AVOZ1-010-914 of the
ASCR. The financial support provided by the Grant Agency of the
ASCR (Grant No. A1010214), by the Grant Agency of the Czech
Republic (202/00/0122), and by RTN project No. HPRN-CT-2000-00143
the EC acknowledged.


\end{document}